\documentclass[fleqn,10pt]{wlscirep}
\title{Months-long real-time generation of a time scale based on an optical clock}

\author[1]{Hidekazu Hachisu}
\author[1]{Fumimaru Nakagawa}
\author[1]{Yuko Hanado}
\author[1,*]{Tetsuya Ido}
\affil[1]{National Institute of Information and Communications Technology,
4-2-1 Nukui-kitamachi, Koganei, Tokyo 184-8795, Japan}
\affil[*]{ido@nict.go.jp}

\begin{abstract}
Time scales consistently provide precise time stamps and time intervals by combining atomic frequency standards with a reliable local oscillator. Optical frequency standards, however, have not been applied to the generation of time scales, although they provide superb accuracy and stability these days. Here, by steering an oscillator frequency based on the intermittent operation of a $^{87}$Sr optical lattice clock, we realized an ``optically steered'' time scale TA(Sr) that was continuously generated for half a year. The resultant time scale was as stable as International Atomic Time (TAI) with its accuracy at the $10^{-16}$ level. We also compared the time scale with TT(BIPM16). TT(BIPM) is computed in deferred time each January based on a weighted average of the evaluations of the frequency of TAI using primary and secondary frequency standards. The variation of the time difference TA(Sr) $-$ TT(BIPM16) was 0.79 ns after 5 months, suggesting the compatibility of using optical clocks for time scale generation. The steady signal also demonstrated the capability to evaluate one-month mean scale intervals of TAI over all six months with comparable uncertainties to those of primary frequency standards (PFSs).
\end{abstract}
\begin{document}

\flushbottom
\maketitle
\thispagestyle{empty}

%\noindent Please note: Abbreviations should be introduced at the first mention in the main text – no abbreviations lists. Suggested structure of main text (not enforced) is provided below.

\section*{Introduction}
Time scales form a foundation of modern society, enabling technologies such as global navigation satellite systems, radio-wave interferometer arrays, and the synchronization of grids for telecommunication networks and electric power systems. In general, stable frequency standards become an infrastructure when signals are integrated to a time scale, by which the benefit of atomic frequency standards is always available as timestamps or scale intervals. While various time scales have been coordinated and operated\cite{BIPM:Annual}, most national standard times are adjusted to Coordinated Universal Time (UTC) since the availability of UTC is solid due to the sufficient redundancy of the atomic clocks involved. The International Bureau of Weights and Measures (BIPM) receives information regarding more than 400 atomic clocks operated in national metrology institutes  and other public laboratories, and then subsequently computes their weighted average\cite{Panfilo:Algo,Panfilo:weight}. The frequency offset of this mean free time scale with respect to the SI second is estimated using primary frequency standards (PFSs) or secondary frequency standards. This calibrated atomic mean time scale is called International Atomic Time (TAI). UTC is the product after adding a phase step of integer seconds to TAI to synchronize the time scale to Earth’s rotation \cite{ITU}. It is noteworthy that UTC is one of the so-called paper clocks, meaning a virtual product computed with a latency of 15–40 days and that the accuracy of its scale interval relative to the SI second lies at the $10^{-16}$ level. Furthermore, the computation results are provided at limited instances at 0 h UTC on every fifth day. BIPM also computes UTCr, which is computed every day and made available every week. The sparse availability of UTC with latency requires national time and frequency laboratories to generate their own time scales in real time for their national standard time. Such a local time scale adjusted to UTC is called UTC($k$), where $k$ stands for laboratory $k$ \cite{BIPM:Annual}. There are various recipes for the generation of UTC($k$). The most stable one is currently realized by steering the frequency of a hydrogen maser (HM) with respect to a local Cs (or Rb) fountain standard \cite{Bauch:UTC(PTB),Rovera:UTC(OP),Russia:CsF,Peil:USNO}.

Microwave frequency standards, which presently realize all atomic time scales, have made steady progress for more than fifty years since the SI second was defined by the Cs hyperfine transition. However, advances in optical frequency standards (OFSs) have accelerated in the last twenty years, resulting in greatly improved stability and accuracy over microwave frequency standards \cite{Ludlow:review}. This situation has led the community of time and frequency standards to seriously consider the future redefinition of the SI second, where the SI second is realized by OFSs and the accuracy of the scale interval in TAI is maintained by OFSs \cite{Riehle}.

This trend to introduce OFSs may also activate attempts to realize more stable UTC(k) with more accurate scale intervals. UTC($k$) needs to be a real-time signal without interruptions. The simplest way to achieve this could be to use an OFS directly as a signal source of UTC($k$). However, a more realistic way being currently discussed is steering a local oscillator with reference to an OFS instead of a microwave frequency standard.
%combined method. While the continuity of a time scale is maintained by an oscillator, which is often called a “flywheel”, the flywheel frequency is measured using a frequency standard, and accordingly, the flywheel frequency is adjusted. Several institutes including NICT currently adopt this method with both the flywheel and calibrator in the microwave regime \cite{JST2008}.
The feasibility of this method with an OFS was investigated by post-processing using data of OFS operations over $10-80$ days\cite{NICT:JPCS,PTB:Optica}.In particular, it was reported by Grebing {\it et al.}\cite{PTB:Optica} that they made a simulation and predicted that the nearly continuous operation of an OFS in future may realize a time scale which is superior to those provided by a state-of-the-art PFS.

Here, in this article, we propose and demonstrate an OFS-based time scale, which only requires temporally sparse operation of an OFS. Together with the operation of a stable HM as a flywheel, an accurate ${}^{87}{\rm Sr}$ lattice clock for the measurement of the maser frequency was operated for $10^4$ s approximately once a week, and the frequency of the flywheel was continuously steered for half a year with reference to the calibrations, leading to a real-time signal of a stable time scale with accurate scale intervals. Our intermittent operations of the OFS also enabled the one-month mean of the TAI scale interval to be estimated uninterruptedly throughout the half year. Evaluations by using an optical clock with an up-time ratio below 2\% make it much easier for OFSs to contribute to TAI, suggesting a novel approach to maintain the accuracy of the TAI scale interval. For this goal, we also discuss the requirements of the OFS operation rate and HM instability, which should be useful for applying the proposed scheme to other combinations of OFSs and flywheels.

\section*{Results}
\subsection*{Architecture of the time scales steered by an intermittently available OFS}
Regardless of the rapid progress in OFSs, the continuous operation of OFSs has not become an easy task owing to the difficulty in maintaining the whole complicated system at its optimum state. Thus, we need to investigate what level of continuous operation is required for an OFS to accurately steer the flywheel frequency. A clue to addressing this issue is found in Fig. 1, where the frequency instabilities of various oscillators including those used in this work are shown. HMs are commonly used as a flywheel owing to their excellent balance between short- and long-term frequency stabilities. The instabilities of HMs are normally lowest at around $10^4-10^5$ s in terms of the Allan deviation. This is due to the existence of a linear frequency drift, and such characteristics are found in Fig. 1 from the instabilities of HM1 and HM2 (red and blue) that we used in this work.  The red curve shows that the HM1 frequency reaches an instability of around $3\times 10^{-16}$. To keep the stability at the $10^{-16}$ level, the frequency of HM1 must be steered to an external stable frequency reference. Cs fountain frequency standards play the role of the calibrator owing to their stable and accurate frequency in the long term, although they require a long averaging time for high precision.  In the case of  conventional Cs fountains using an HM as a reference\cite{Kuma}, it takes more than a week to evaluate a frequency with a statistical uncertainty of $5\times 10^{-16}$ (orange line in Fig. 1). Even for a more stable Cs fountain employing a cryogenic oscillator \cite{SYRTECsF} or photonically generated microwave \cite{OpticalMW} as a reference (green line), operation for over $3\times 10^4$ s is still necessary \cite{CsFFiberLink}.

The instability of OFSs, on the other hand, is much lower than that of microwave frequency standards. Since the instability of the Sr optical lattice clock employed here (purple) is an order of magnitude lower than that of other microwave standards, the instability in measuring the frequency of an HM using an OFS is determined solely by the HM. Overall, the plots in Fig. 1 indicate that the operation of an OFS for only $10^4$ s is required to evaluate the HM frequency with accuracy at the mid-$10^{-16}$ level. This is much shorter than for the case of using Cs fountains.
In longer time scale, the temporal variation of an HM frequency comprises a linear frequency drift and a stochastic phase fluctuation \cite{HM,NICT:JJAP}. Thus, the prediction of the linear drift from the latest records of the HM frequency makes it possible to cancel the frequency drift to a large extent. Then, the residual noise, which is expressed by the three-sample variance (Hadamard variance) $\sigma_{\rm H}^2\left( \tau \right)$, forms a flicker noise floor that extends to rather long averaging time. This extended flicker floor allows the time scale to rely on the HM for a certain amount of time. OFSs are highly advantageous in this regard because their high stability enables the evaluation of the HM frequency in a short averaging time. We examined the feasibility of this strategy in our case considering the instability of available HMs. The Hadamard deviations of ${\rm HM1 - HM2}$ divided by $\sqrt{2}$, UTC-HM1, and UTC-HM2 are shown in Fig. 1 as dashed curves, indicating flicker floors of HM1 and HM2 at the $(3 - 5) \times 10^{-16}$ extending to an averaging time of one week or more. Thus, it is feasible to compose a time scale with instability at the $10^{-16}$ level by the intermittent operation of an OFS once a week.
\begin{figure}[ht]
\centering
\includegraphics[width=.8\linewidth]{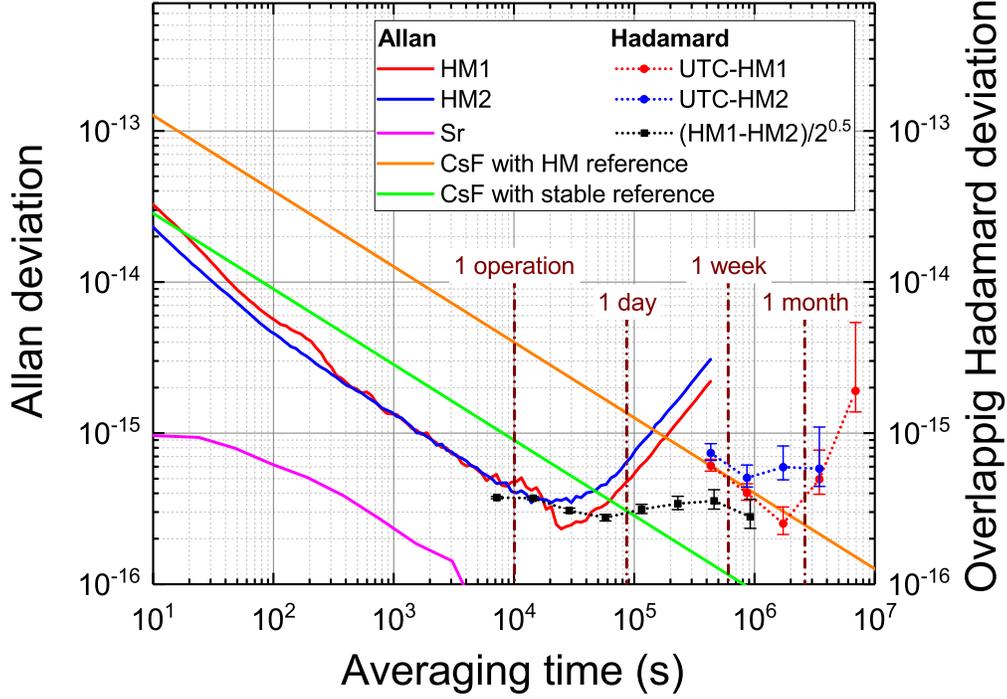}
\caption{{\bf Frequency instabilities of various oscillators, including microwave and optical frequency standards as well as the two HMs and Sr lattice clock used in this work.} The much lower instability of the lattice clock allows the HM mean frequencies to be evaluated in a shorter time. The dashed curves are relevant overlapping Hadamard deviations. The deviation of HM1 vs HM2 divided by $\sqrt{2}$ is shown as black squares for short averaging times of less than 10 days. The deviations of HM1 and HM2 for longer averaging times are calculated against UTC.}
\label{fig:Stabilities}
\end{figure}

\subsection*{Time scale generation by steering a maser frequency with respect to an optical clock}

Our system for the generation and evaluation of an optically steered time scale is depicted in Fig. 2. For the generation, an HM (HM1) and a $^{87}{\rm Sr}$ lattice clock (NICT-Sr1) act as a flywheel and a calibrator, respectively. As shown in Fig. 1, HM1 has instability at the $4\times 10^{-16}$ level at an averaging time of $1\times 10^4$ s, from which the Hadamard deviation is extended to a flicker floor of around $3\times 10^{-16}$ up to three weeks. Thus, we operate NICT-Sr1 for only $10^4$ s per operation, after which the time scale relies on HM1 until the next OFS operation. NICT-Sr1 as a calibrator has a systematic uncertainty at the $10^{-17}$ level \cite{NICT:OE}, which proves the validity of utilizing it as an accurate frequency reference. The steering was based on the absolute frequency, which we determined in Ref. \citenum{NICT:APB}.

The offset and drift in HM1 frequency were corrected using a phase micro stepper (PMS). With reference to HM1, the PMS generates a frequency shifted from HM1 by a small amount. The linearly drifting HM1 frequency between two OFS operations was estimated from the time series of the HM1 frequencies evaluated by NICT-Sr1 in the past. Then, the corresponding frequency offset at the moment was repeatedly estimated and added to the PMS with the opposite sign, resulting in a drift-free signal, TA(Sr). Thus, TA(Sr) was generated independently without any reference time scales.

\begin{figure}[ht]
\centering
\includegraphics[width=.85\linewidth]{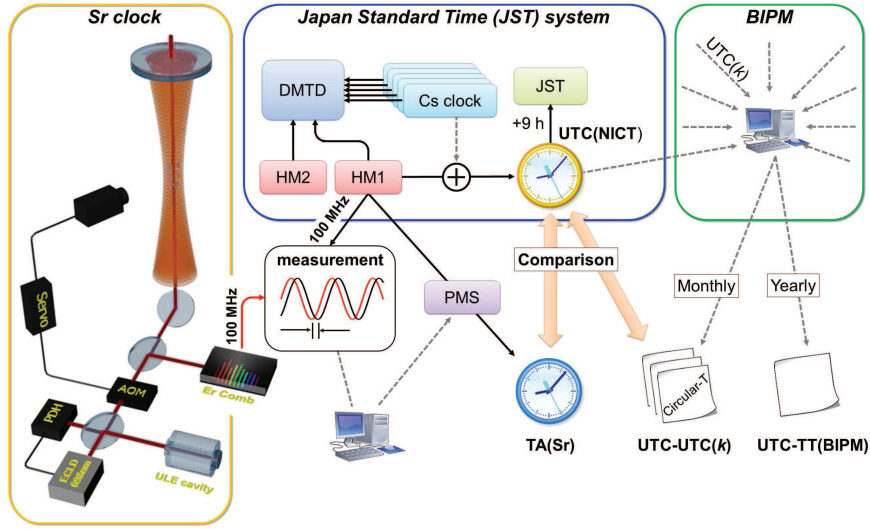}
\caption{{\bf Schematic diagram of the realization and evaluation of an optically steered time scale.} A strontium lattice clock and a frequency comb stabilized to the clock laser generate a microwave with a frequency of exactly 100 MHz. We operate the lattice clock once a week or more frequently, and record the differential phase between HM1 and the OFS-based microwave at the start and end of the operation, from which the mean fractional frequency of HM1 is calculated. The results obtained in the past 25 days allow us to predict the linear drift of the HM1 frequency. A phase micro stepper (PMS) externally shifts the HM1 frequency slightly so as to cancel the predicted linear drift, resulting in the time scale TA(Sr). TA(Sr) and UTC are mediated by UTC(NICT). Thus, the accuracy and stability of TA(Sr) were evaluated by obtaining the time difference from the virtual “paper clocks” UTC and TT(BIPM16). DMTD : dual mixer time difference system\cite{NICT:DMTD}, JST : Japan Standard Time.}
\label{fig:schematicDiagram}
\end{figure}

The free-running HM1 signal at 100 MHz was compared with the 100 MHz signal downconverted from NICT-Sr1 by recording the phase difference. The mean frequency of HM1 with respect to the OFS-based reference was calculated from the differential phase between the start and end of the OFS operation. Thus, it is convenient to introduce the fractional frequency of HM1 $y_i$ for the $i$th OFS operation as

\begin{equation}
y_i=\frac{f_{\rm HM1}}{f_{\rm OFS}}-1=\frac{\phi_{\rm F}-\phi_{\rm I}}{\Delta_{\rm op}},
\end{equation}
where $f_{\rm HM1}, f_{\rm OFS}, \phi_{\rm F}, \phi_{\rm I},$ and $\Delta_{\rm op}$ are the HM1 frequency, the OFS-based reference frequency, the relative phases at the end and start, and the duration of the OFS operation, respectively. Note that $\phi_{\rm F}$ and $\phi_{\rm I}$ are in the unit of seconds, resulting in $y_i$ being a fractional quantity.

On the basis of $y_i (i=0,\ldots ,N)$ obtained in the past time interval $T$, the linear drift of $y(t)$ was predicted as shown in Fig. 3(a). Here, $t_i$ and $t_{\rm S}$ are the time of the $i$th operation of the OFS and the time when the prediction of the linear drift is updated, respectively. It is assumed here that the OFS was operated $N+1$ times inside the drift evaluation interval of $t_{\rm S}-T<t<t_{\rm S}$. The latency for incorporating new measurement results is less than 18 h, which is negligible compared with other parameters such as $T$ and $t_{i+1}-t_i$.  At time $t_{\rm S}$ which occurs every 4 h regardless of OFS operations, linear fitting to $y_i \left( i=0, \ldots , N; t_{\rm S}-T<t_i<t_{\rm S} \right)$ was performed, resulting in the estimation $\hat{y}\left( t\right)$ of the HM fractional frequency $y(t)$ being

\begin{equation}
\hat{y}\left( t\right) = \overline{y(t_{\rm S})}+d\left( t_{\rm S}\right) \cdot t,
\end{equation}
where $\overline{y(t_{\rm S})}$ and $d(t_{\rm S})$ are the offset and drift rate of the fitting predicted at $t_{\rm S}$, respectively. Note that $\hat{y}(t)$ changes not only immediately after the OFS operation but also when the earliest evaluation $y_0$ becomes beyond the drift evaluation interval $t_{\rm S}-T < t < t_{\rm S}$. We need to consider a tradeoff in the determination of $T$. More data points in a longer $T$ may reduce the statistical error in the linear fitting, whereas the fitting over a long duration suffers from the error due to the quadratic and higher order terms of $y(t)$. We set $T$ to 25 days, until which $y(t)$ is well approximated as linear because the flicker floor of the Hadamard deviation $\sigma_{\rm H}(\tau)$ of HM1 remains for up to 20 days (Fig. 1). The OFS operation was performed once a week or more frequently to obtain four or more points for the fitting. This corresponds to $N=3$ or more. Soon after $\hat{y}(t)$ is renewed at $t_{\rm S}$, we renewed the steering frequency at PMS $\Delta y_{\rm PMS}(t)$, to cancel $\hat{y}(t)$. Considering the potential noise in changing $\Delta y_{\rm PMS}(t)$, the update rate of $\Delta y_{\rm PMS}(t)$ was suppressed to once every 4 h.

We began the intermittent operation of the OFS (NICT-Sr1) in April 2016. After one month of preliminary operations for the initial characterization of HM1, we started the steering on May 1st, 2016 (MJD 57509), and continued it for five months to the end of September (MJD 57661). The red points in Fig. 3b show the measured fractional frequency $y_i$ of HM1 relative to NICT-Sr1 over half a year. The results clearly show the stable linear drift of the HM1 frequency. The details of the evaluation in Fig. 3(b) are shown in Fig. 3c, where the differences from the linear fitting line (green dashed line) in Fig.3b are shown. The purple line in Fig. 3c shows $-\Delta y_{\rm PMS}(t)$. %Soon after NICT-Sr1 is operated or the earliest OFS operation falls off from the drift estimation interval, the purple line suddenly changes, indicating the update of the prediction.
The purple line has a fine sawlike structure with a period of 4 h since the frequency correction at the PMS (blue line in Fig. 3(a)) is constant for 4 h. The deviations of the red points from the purple line, which are the difference in the prediction $\hat{y}(t_N)$ from $y_N$, are shown in the lower panel of Fig. 3c. In most cases, they are lower than $1\times 10^{-15}$.

\subsection*{Comparison with reference time scales}
For the characterization of TA(Sr), we need other independent reference time scales for comparison. We chose UTC and TT(BIPM) as references since they are linked to UTC(NICT). UTC(NICT) is an atomic time scale that is used as the origin of Japan Standard Time (JST). The phases of HM1, HM2, and TA(Sr) with respect to UTC(NICT) are continuously monitored by a precise phase measurement system (DMTD, dual mixer time difference system)\cite{NICT:DMTD} contained in the JST system \cite{JST2008}. The time difference of UTC(NICT) relative to other UTC($k$) is routinely measured by satellite-based precision time and frequency transfer methods \cite{BIPM:Annual}. By incorporating such data sent from public laboratories worldwide, BIPM publishes UTC(NICT)-UTC and its uncertainty in the monthly report Circular T.

The blue points in Fig. 4 show the time difference ${\rm TA(Sr) - UTC}$. While TA(Sr) is a real-time signal, UTC is virtual and available only every 5 days. The time offset was initialized to zero on May 1st, 2016 (MJD 57509). TA(Sr) gradually became delayed relative to UTC and the difference reached 8.3 ns after five months. This monotonic change corresponds to a mean frequency bias of $6\times 10^{-16}$. This, however, does not indicate that the scale interval of TA(Sr) is different from that of the SI second by that amount. Instead, this deviation is caused by the fact that the scale interval of UTC was smaller than that of the SI second. The magnitude of the deviation is suggested in Circular T, where the scale interval of UTC is estimated by considering the calibration using the results reported from PFSs worldwide. Circular T Nos. 341–345 indicate that the scale interval of UTC was shorter than the SI second by a fractional amount at the mid-$10^{-16}$ level during our study.

\begin{figure}[hbt]
\centering
\includegraphics[width=.75\linewidth]{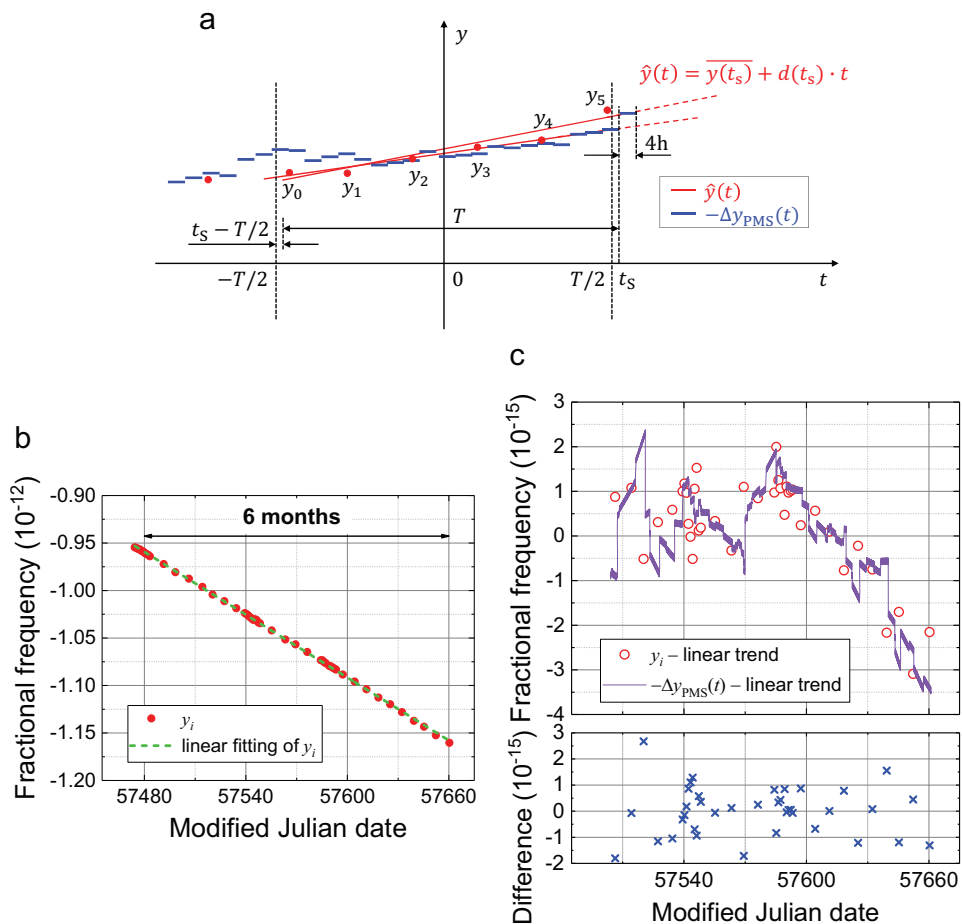}
\caption{{\bf Scheme of frequency steering. a,} As shown by the red dashed line, the HM frequency $y(t)$ was predicted by extrapolation of a linear fitting to past HM frequencies $y_i (i=0\ldots N)$, which were obtained in the latest duration of $T$. Here, the case of $N=5$ is drawn. The steering frequency $\Delta y_{\rm PMS}(t)$ was changed every 4 h. We determined the value of $\Delta y_{\rm PMS} (t)$ entered in the PMS so that the time integration of $\hat{y}(t)+\Delta y_{\rm PMS}(t)$ becomes null; in other words, $-\Delta y_{\rm PMS}(t)$ and $\hat{y}(t)$ intersect at the middle of one step. Note that the number of 4 h intervals between $y_i$ and $y_{i+1}$ shown here is much smaller than that of the real situation since two calibrations are typically separated for one week. {\bf b,} Change in the frequency of a hydrogen maser (HM1) evaluated using the $^{87}\rm Sr$ lattice clock. Stable linear drift was found. {\bf c,} The red points in {\bf{b}} are shown in the upper panel as the difference from the linear fitting (green dashed line in {\bf{b}}). The magnitude of the frequency shift subtracted at the PMS is also shown in purple. The purple curve changes soon after a new red point updates the prediction of the HM frequency, or at the instants when the earliest red point falls off from the drift evaluation interval. The lower panel shows the difference in $\hat{y}(t_N)$ from $y_N$.}
\label{fig:Steering}
\end{figure}

We can discuss this point quantitatively by comparison with the time scale, TT(BIPMXY) \cite{BIPM:TT(BIPM)}. The scale interval of TT(BIPMXY) is highly accurate since it is computed annually by a postprocessing that incorporates all calibrations provided by primary or secondary frequency standards worldwide. The “XY” denotes the last year of the computation interval. The most recent time scale, TT(BIPM16), was issued in January 2017, and the relative difference TT(BIPM16) ‒ UTC is drawn as a black dashed line in Fig. 4 with the offset nulled on May 1st. It is clear that TA(Sr) shows good agreement with TT(BIPM16). The difference TA(Sr) ‒ TT(BIPM16) is drawn as red points and was 0.79 ns after five months’ generation.

The superb stability of TA(Sr) is observed as the overlapping Allan deviation of TA(Sr) ‒ UTC shown in the inset. Note that this instability includes the fluctuation of the time link between UTC and UTC(NICT). This extra uncertainty is hidden in the instability between TA(Sr) and UTC. As we discussed in Ref. \citenum{NICT:OE}, the link instability appears for up to an averaging time of 10 days. For an averaging time of 20 days, however, the link uncertainty is reduced to $2\times 10^{-16}$, and the instability of $3.9\times 10^{-16}$ at 20 days shown in the inset is predominantly determined by the instability of TA(Sr) and UTC. Considering that UTC has an instability of $3\times 10^{-16}$ at one month \cite{BIPM:Annual}, the residual instability of $1.5\times 10^{-16}$ could be attributed to the instability of TA(Sr), demonstrating the instability of TA(Sr) that rivals state-of-the-art paper clocks.

\begin{figure}[ht]
\centering
\includegraphics[width=.8\linewidth]{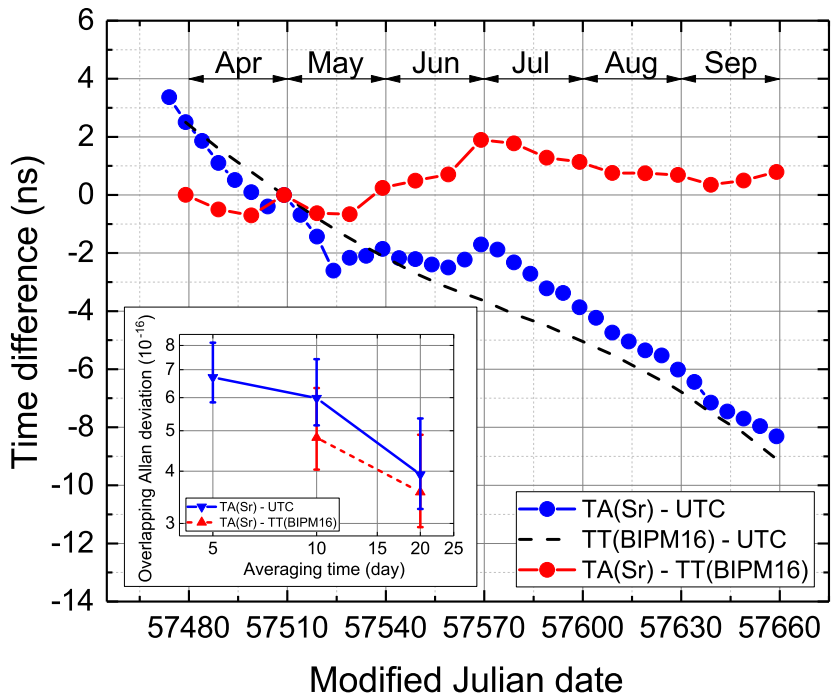}
\caption{{\bf Behaviors of the realized time scale TA(Sr) and TT(BIPM16) with respect to UTC (blue filled circles and dashed line, respectively).} TA(Sr) is consistent with TT(BIPM16). The red filled circles show the difference ${\rm TA(Sr)}-{\rm TT(BIPM16)}$, showing the agreement at the sub-ns level after five months. The inset shows the corresponding overlapping Allan deviation of ${\rm TA(Sr)}-{\rm UTC}$ and ${\rm TA(Sr)}-{\rm TT(BIPM16)}$.}
\label{fig:vsUTC}
\end{figure}

\subsection*{Requirement for the HM stability and OFS operation rate }
For the time scale steered by the intermittent operation of an OFS, it is valuable to investigate how often we need to operate optical clocks depending on the stability of the flywheel. The uncertainty of the time scale is predominantly attributed to the prediction error of the linear trend as well as the stochastic phase of the flywheel. It is possible to calculate these uncertainties analytically by considering the error of least-squares fitting as well as the noise characteristics of the flywheel. The estimation, whose details are described in ``Methods'', results in the deviation of the phase accumulated between two OFS operations being
\begin{equation}
E\left[ \left|\Delta \phi\right| \right]=\left( \epsilon_{\rm p}^2+\epsilon_{\rm F}^2\right) ^{1/2} = \frac{T}{N}\left[\frac{(2N+1)(2N+3)}{N(N+1)(N+2)}\sigma_{\rm p}^2+\frac{1}{\ln 2}\sigma_{\rm F}^2 \right]^{1/2},
\end{equation}
where it is assumed that $N+1$ operations are homogeneously distributed in $T$, $\sigma_{\rm p}$ is the statistical uncertainty in the evaluation of the flywheel frequency by OFS operation, and $\sigma_{\rm F}$ is the magnitude of the flicker floor in the Hadamard deviation of the flywheel. The contributions originating from $\sigma_{\rm p}$ and $\sigma_{\rm F}$ to the total phase error are denoted as $\epsilon_{\rm p}$ and $\epsilon_{\rm F}$, respectively.

The dependence of $E\left[ |\Delta \phi|\right]$ on $N$ is shown in Fig. 5 as black squares along with $\epsilon_{\rm p}$ and $\epsilon_{\rm F}$. Here, the parameters $\sigma_{\rm p}$ and $\sigma_{\rm F}$ are $4\times 10^{-16}$ and $3\times 10^{-16}$, respectively, and we consider the case of $T = 30$ days for generality. Thus, $N$ corresponds to the number of calibrations per month. $E[|\Delta \phi|] N^{1/2}$, namely the red points in Fig. 5, is the integrated extra phase in one month for the ideal case that $\Delta \phi$is uncorrelated with the adjacent free evolution. This hypothesis is introduced as a simplification, although the detrended HM shows not white frequency noise but flicker frequency noise. Nevertheless, we can analytically estimate the possible phase error owing to this simplification, and we consider that $E[|\Delta \phi|] N^{1/2}$ can still be used as a guide to choose the parameters relevant to the intermittent optical steering. The estimation of the phase error integrated over five months, namely $E[|\Delta \phi |] (5N)^{1/2}$, is also shown as blue points to compare with our result. Here, the case of $N=4$, namely, OFS operations once a week, yields a time difference of 1.5 ns. This level is similar to the final difference of 0.79 ns obtained in our experiment as shown in Fig. 4. Figure 5 also shows that the extra phase in one month is 0.66 ns in the case of the once-a-week operation. Taking into account the fact that the notification of UTC by Circular T has a mean latency of one month, it seems that maintaining the synchronization of UTC($k$) within 1 ns is feasible for the optically steered time scale.

\begin{figure}[ht]
\centering
\includegraphics[width=.8\linewidth]{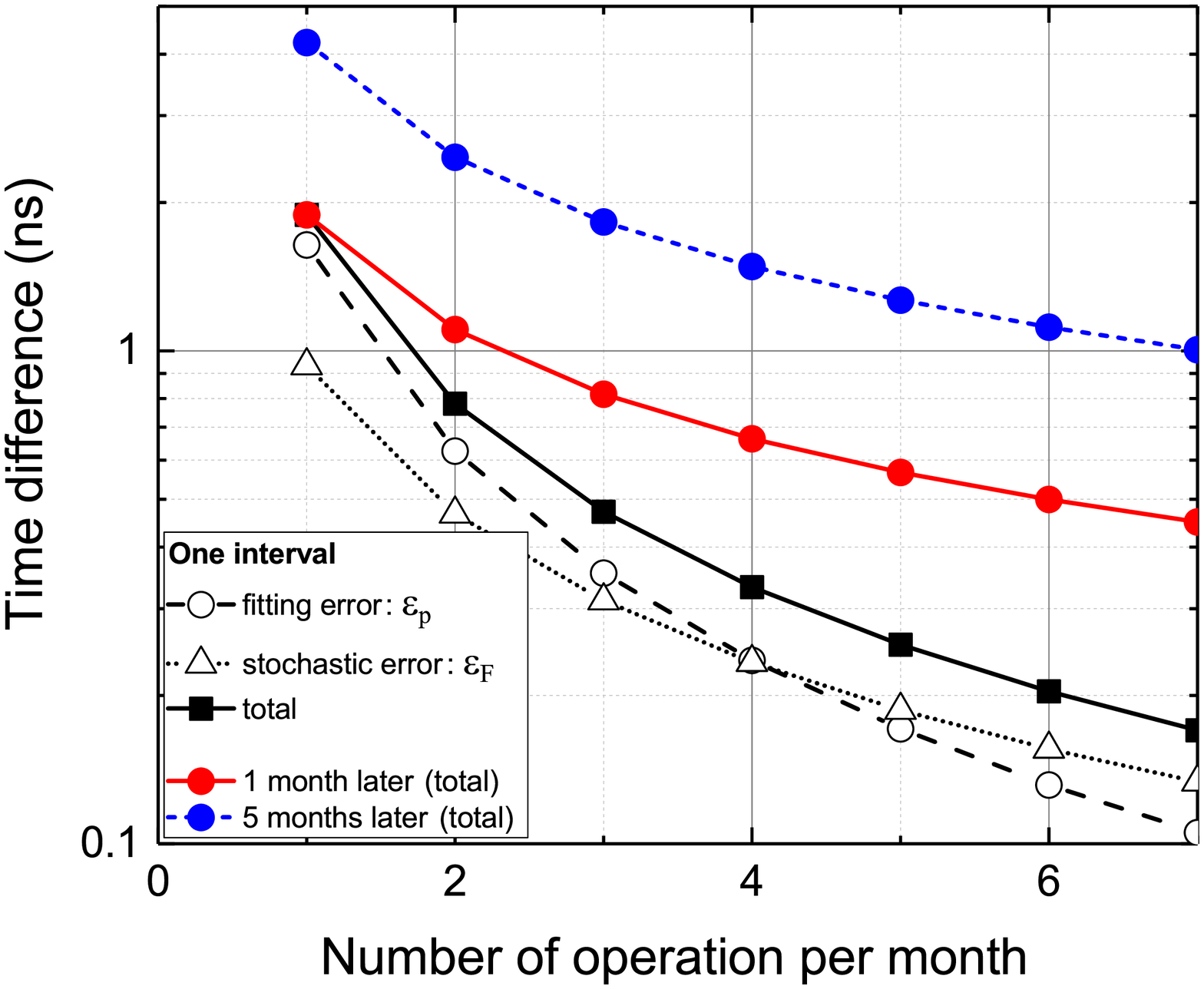}
\caption{{\bf Errors in an intermittently steered time scale as a function of the number of OFS operations per month.} The black filled squares are the errors accumulated in one interval between two OFS operations. The error is attributed to the fitting error in predicting the linear frequency drift (open circles) as well as the stochastic phase of an HM (open triangles). The red and blue points are estimations of the accumulated total error in one month and five months, respectively. Note that these are analytically estimated using equation (3).}
\label{fig:SimpleCalc}
\end{figure}

\subsection*{Simulation for the case of another noisier HM and infrequent OFS operation}
The dependences of the time scale on parameters such as HM stability and OFS operation rate were investigated by simulations based on the actual record of the phase in HM1 as well as another HM (HM2) with a higher flicker floor level of $5\times 10^{-16}$. The JST system contains both HM1 and HM2, and records their relative phase with respect to UTC(NICT) every second. Using this record, we can extract the mean frequency ratio of the two HMs over an OFS operation time of $10^4$ s, enabling postprocessing evaluations of the HM2 frequency. Then, each point in Fig. 3b is converted to the value for HM2, with which we can determine the adjustment frequency at a virtual PMS used for HM2, enabling the simulation of the case that HM2 is used as the flywheel.

Furthermore, it is possible in simulations to partly omit the steering to investigate the dependence of the instability on the OFS operation rate. It is better in principle to perform OFS operations with a temporally homogeneous distribution as an HM has flicker noise in the Hadamard deviation. All points in a lumped evaluation may lie on one side of the linear trend, which may result in the  incorrect prediction of the linear frequency drift. Considering these issues, we picked the OFS operations that realize a rather homogeneous distribution of once a week or once every two weeks, and made simulations for the cases of HM1 and HM2 as the flywheel oscillator. These cases correspond to the parameters of $T=21$ days and $N=3$ for the once-a-week calibration, and $T=28$ days, $N=2$ for the calibration every two weeks. Note that it is possible to obtain two curves for the case of evaluations once every two weeks by choosing odd and even weeks.

The results of six simulations (HM1 and HM2 every week and on odd and even weeks) were compared with that of TT(BIPM16) as shown in Fig. 6, where the expected phase excursions derived from the simple estimation discussed above are also shown as shaded areas. Here, we ignored the UTC-UTC(NICT) link error since it is negligibly small for an averaging time of more than 20 days. It is clear that the time scales using HM1 (solid curves) are more stable than those using HM2 (dashed curves), reflecting the difference in the flicker floor. The cases of infrequent evaluations once every two weeks are drawn as green and orange curves, denoting the choice of odd and even weeks, respectively. Having the half rate of steering causes larger phase excursions for the case of HM2. We can derive the instabilities from the simulation results shown in Fig. 6. The overlapping Allan deviation over 40 days is around $3\times 10^{-16}$ for the two operation rates using HM1. The deviations using HM2 with the calibration rates of once a week and once every two weeks were 5 and 7 parts per $10^{16}$, respectively.

\begin{figure}[ht]
\centering
\includegraphics[width=.8\linewidth]{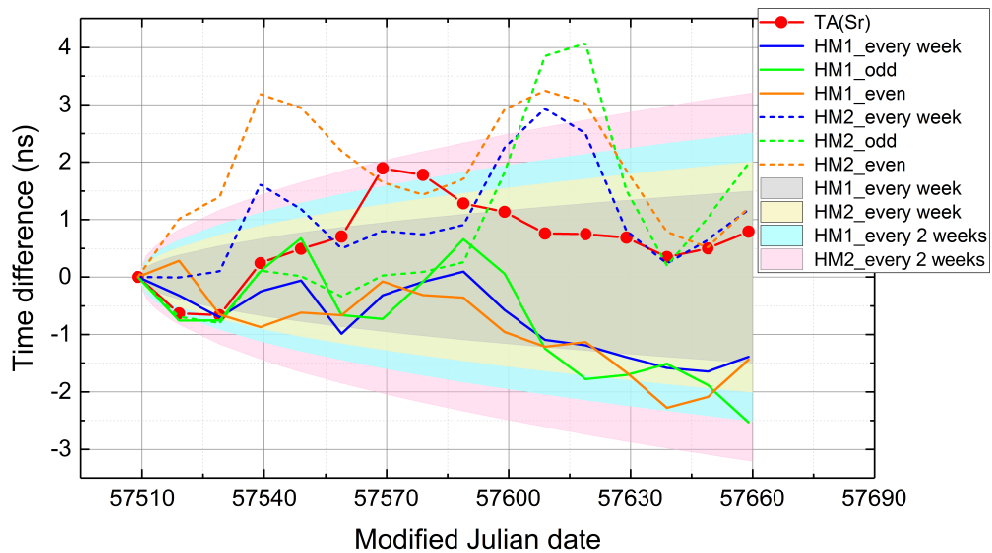}
\caption{{\bf Time differences from TT(BIPM16) simulated using actual record of HM behaviors.} The realized time scale TA(Sr) is also shown for reference as red points. The blue solid curve is the simulation result when the flywheel frequency was evaluated using the OFS almost regularly once a week. The green and orange curves show the cases that the parameters were steered every two weeks in odd and even weeks, respectively. The less stable HM, HM2, was employed to obtain the three dashed curves. }
\label{fig:Simulation}
\end{figure}

\subsection*{Evaluation of the TAI scale interval}
Once a highly stable time scale is realized, the mean frequency difference between TAI and TA(Sr) in a longer interval is derived using equation (1). Employing a longer interval is critically important to reduce the uncertainty as TAI-TA(Sr) always includes the uncertainty of the time link UTC-UTC(NICT). The one-month mean frequency of UTC is normally calibrated using primary or secondary frequency standards, and the result is published in Circular T as the fractional deviation of the scale interval of TAI from that of the SI second.

The resultant mean frequency difference between TAI and TA(Sr) is shown in Fig. 7 as red open circles, where the difference was converted to a fractional scale interval following the notation in Circular T. It also shows the results of the calibration using other references such as TT(BIPM16) and state-of-the-art PFSs (PTB-CSF2 \cite{PTBCSF2} and SYRTE-FO2 \cite{SYRTECsF}); the estimation of the uncertainty for the result of TA(Sr) is not straightforward since the uncertainty depends on the number of OFS operations as well as on how homogeneously the measurements are distributed in time \cite{NICT:JJAP,Yu}.

We therefore employed another method of TAI calibration that allows an easier estimation of the uncertainty. This method does not use all the results of OFS operations but selects five or six data for the month that are separated by nearly one week, and the linear drift in the fractional mean frequency of the HM is estimated by linear fitting using those five or six points. We can evaluate the uncertainty in this case by assuming that the intermittent measurements are homogeneously distributed over one month. Furthermore, as a real-time signal is not required for the TAI calibration, it is possible to use the mean phase of multiple HMs to mitigate the effect of a sporadic phase excursion of a specific HM \cite{NICT:OE}. Table 1 shows a typical uncertainty budget in this proposed scheme of the TAI calibration. The uncertainties due to the extra phase of HM in the OFS down time are dominant contributors. The details of the contributions are described in ``Methods''. The total uncertainty without the uncertainty of standard frequency amounted to $3.6\times 10^{-16}$, which was at a level similar to those in the evaluation using the state-of-the-art PFS. The absolute frequency of the $^{87}{\rm Sr}$ clock transition has been measured in various laboratories over the years, where details including references are given in ``Methods''. The weighted mean of all absolute frequency measurements is $\nu_{\rm mean} = 429\ 228\ 004\ 229\ 873.04$ Hz with a fractional uncertainty of $9.1\times 10^{-17}$. Here, we adopted this $\nu_{\rm mean}$ as the standard frequency, which leads to the total uncertainty of the TAI calibration being $3.7\times 10^{-16}$.

The result of this TAI calibration using NICT-Sr1 combined with the mean of two HMs (HM1 and HM2) is depicted in Fig. 7 as red filled circles. Here, we adopted the mean as the standard frequency. It shows that NICT-Sr1 was able to derive the scale interval of TAI over the half year. The results of NICT-Sr1 are consistent with the calibrations provided by the other two PFSs. This consistency is quantitatively investigated as follows. First, we calculated the mean of the three calibration results provided by the two Cs fountains and NICT-Sr1. Then, we obtained the mean deviations of the calibrations from this mean for the three references, which were $6.3\times 10^{-16}, 7.8\times 10^{-16}$, and $2.6\times 10^{-16}$ for PTB-CSF2, SYRTE-FO2, and NICT-Sr1, respectively. The deviation of the NICT-Sr1 calibration is smaller than those of the two PFSs, indicating that the evaluation using Sr is consistent with those using other PFSs. In addition, the deviations of the calibrations by PTB-CSF2, SYRTE-FO2, and NICT-Sr1 with respect to the calibration by TT(BIPM16) were also calculated to be $2.3\times 10^{-16}, 2.8\times 10^{-16}$, and $2.8\times 10^{-16}$, respectively. Note that TT(BIPM16) is effectively the average of all calibrations by PFSs, among which PTB-CSF2 and SYRTE-FO2 are the main contributors, and consequently, TT(BIPM16) and the two fountains have strong correlations. In spite of the correlations, the calibration by NICT-Sr1 is not significantly different from that by TT(BIPM16). These two investigations demonstrated that the intermittent operation of an OFS can be used for calibrating TAI with a similar level of uncertainty to that regularly provided by PFSs.

\begin{figure}[ht]
\centering
\includegraphics[width=.8\linewidth]{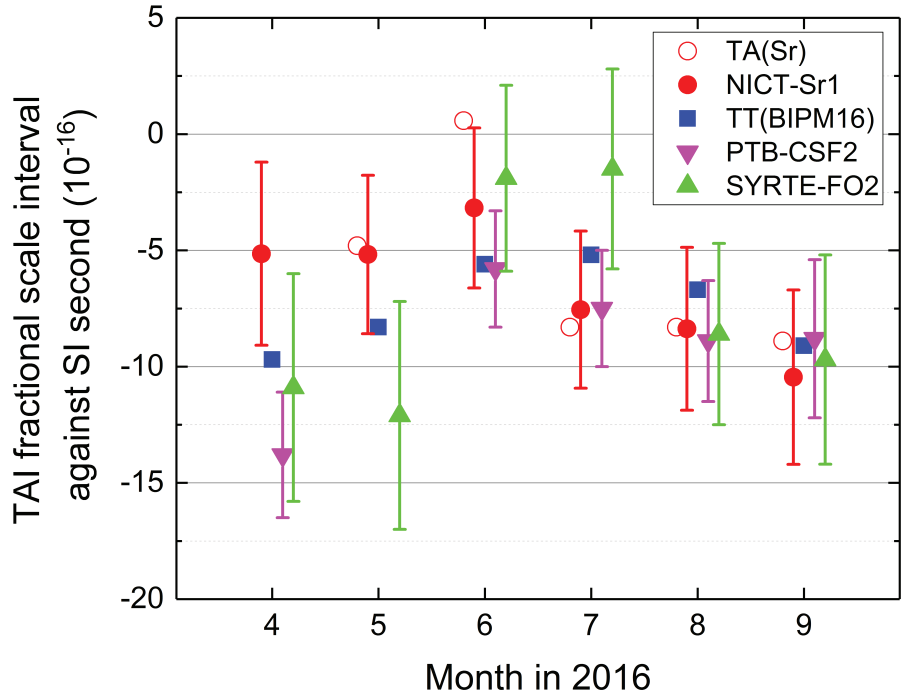}
\caption{{\bf Evaluation of one-month-mean TAI scale interval.} The red open circles are calculated using the time scale TA(Sr), where the change of the time difference ${\rm UTC}-{\rm TA(Sr)}$in one month is divided by the duration of one month. The results obtained with the intermittent operation of the Sr lattice clock (red filled circles, see text) are consistent with those of other methods such as state-of-the-art PFSs (PTB-CSF2 and SYRTE-FO2).}
\label{fig:TAICalib}
\end{figure}

\begin{table}[ht]
\centering
\begin{tabular}{lr}
{\bf Sr atomic frequency standard} 	&  \\
statistical $(u_{\rm a})$					& 3 \\
systematic $(u_{\rm b})$					& 7 \\
{\bf Link}\\
$u_{\rm l/Lab}$							&  \\
HM: trend estimation $(u_{\rm l/HMtrend})$	& 26 \\
HM: stochastic phase noise $(u_{\rm l/HMstoch})$ & 18 \\
DMTD system $(u_{\rm l/DMTD})$			& 5 \\
$u_{\rm l/Tai}$						& \\
UTC-UTC(NICT) link					& 14 \\
\hline
{\bf Subtotal} $(u_{\rm sub})$			& 36 \\
Standard frequency $(u_{\rm SDOM})$		& 9 \\
\hline
{\bf Total}							& {\bf 37}
\end{tabular}
\caption{\label{tab:error}{\bf Uncertainty budgets in the evaluation of the TAI scale interval using the intermittent operation of an OFS.} All numbers are in parts of $10^{-17}$. Details of the contributors are described in ``Methods''.}
\end{table}

\section*{Discussion}

State-of-the-art Cs fountains have realized accuracy below $10^{-15}$. While some fountains \cite{SYRTECsF, PTBCSF2} including the Rb fountains in SYRTE \cite{SYRTERbF} and USNO \cite{USNO} are operated almost continuously, fewer new fountains have recently begun their operations. The community of time and frequency standards is instead inclined to develop optical standards. This situation is reflected in the calibration of the TAI scale interval by contributing PFSs, which are reported in Circular T. Smaller number of calibrations have been reported in the last few years, implying the reduced redundancy of PFSs in maintaining TAI at the low $10^{-16}$ level. Ideally, the greatly improved accuracy and stability of OFSs would be reflected in the calibration of TAI. However, the achievable accuracy for TAI calibration is currently limited by the link uncertainty due to the satellite-based time comparisons between a local clock and TAI, requiring a long averaging time. Considering these circumstances, it is productive to generate time scales using the intermittent operation of optical clocks as demonstrated here until improved links, such as intercontinental fiber links, or satellite-based optical frequency comparisons \cite{OpticalFreespaceLink} allow TAI to fully benefit from the superb accuracy and stability of OFSs distributed worldwide. The local oscillator in such a case may also be replaced with an optical oscillator using an optical cavity with low long-term instability. Monocrystalline silicon \cite{PTBSi} is a promising material for such an optical oscillator as it has intrinsically low aging drift. The combination of fiber link networks and OFSs will further increase the reliability of optically steered time scales because various optical standards in separate laboratories linked by fibers can share the task of providing frequent calibrations for the steering.

\section*{Methods}
\subsection*{Sr lattice clock}
The details of NICT-Sr1, the $^{87}{\rm Sr}$ lattice clock operated in this work, are described in Ref. \citenum{NICT:JJAP,NICT:OE}. The Sr atoms are laser-cooled using a two-stage laser cooling technique and loaded to a vertically oriented one-dimensional optical lattice. The total systematic uncertainty of the Sr system was reduced over the course of six months from 8.4 to 5.0 parts per $10^{17}$ predominantly by eliminating stray electric charge on the chamber windows \cite{NICT:OE}. Thus, the systematic uncertainty owing to the OFS is always less than that of the most accurate PFS \cite{NIST-F2}. The major systematic uncertainties in our system are the blackbody radiation shift and lattice light shift. We employed the absolute frequency obtained in Ref. \citenum{NICT:APB} for the steering, which was $5\times 10^{-16}$ lower than that of the CIPM recommendation on that date. This frequency is consistent with those evaluated in other laboratories \cite{PTB:Optica,SYRTESr} with a fractional difference of less than $2\times 10^{-16}$.

The optical frequency at the wavelength of 698 nm as the output of NICT-Sr1 is first downconverted to a microwave frequency using a commercial Er-fiber frequency comb as shown in Fig. 2. The downconverted signal with a frequency of exactly 1 GHz is derived from the fourth harmonic of the repetition rate (= 250 MHz), assuming the absolute frequency of the clock transition obtained in Ref. \citenum{NICT:APB} as described above. Then, the frequency of the 1 GHz microwave is divided tenfold to 100 MHz, and finally, the relative phase difference of the 100 MHz signal of HM1 from this optically generated microwave was recorded every second by a commercial frequency stability measurement set. The two 100 MHz signals are analog-to-digital converted in the instrument with a sampling rate of 64 Ms/s, from which the phase difference is recorded every second. The cumulative change in the phase difference during the measurement leads to a mean fractional frequency difference between the HM and NICT-Sr1. The phase measurement set has a system noise below $1\times 10^{-17}$ over an integration time of $10^4$ s.

\subsection*{Microwave system used to generate time scale and its link to TAI}
The steering of the frequency is implemented using a PMS (Microsemi AOG-110). On the basis of an input signal at 5 MHz provided from HM1, the PMS can generate a signal with a fractional frequency shift of up to $5\times 10^{-8}$. The frequency offset is set with a resolution of $1\times 10^{-19}$, although the instability caused by the system noise of the PMS is $3\times 10^{-13}/\tau$ where $\tau$ is the averaging time in seconds. The 5 MHz signal steered by the Sr clock, namely TA(Sr), is sent to the dual mixer timing difference (DMTD) system contained in the JST system, where the phase difference between TA(Sr) and UTC(NICT) is recorded every second. The system noise of the DMTD system was measured to be $1\times 10^{-16}$ with an averaging time of $10^4$ s \cite{NICT:DMTD}. The GPS data are routinely obtained by feeding UTC(NICT) to a GPS receiver as a reference, by which the difference between UTC(NICT) and GPS time is obtained. This data is sent to BIPM together with time differences between UTC(NICT) and other local atomic clocks such as commercial Cs clocks and HMs. BIPM calculates a weighted mean of atomic clocks (Échelle Atomique Libre, EAL) \cite{Panfilo:Algo} by analyzing such data from various institutes worldwide. Furthermore, BIPM compensates the frequency offset of EAL by incorporating the calibration provided by PFSs operated in NMIs. This correction was constant for more than four years until the end of 2016, which includes the period of the demonstration here. Thus, the deviation of the TAI scale interval relative to the SI second, which amounted to the $10^{-16}$ level as shown in the past Circular T, was left as it was. This deviation is the object calibrated by PFSs and an optical lattice clock here.

BIPM computes another paper clock in annual postprocessing with an accuracy of around 2 parts in $10^{16}$ by employing an algorithm to evaluate the scale interval of TAI with respect to calibration data randomly provided by PFSs\cite{Azoubib}. The resultant time scale is published as the time offset from TAI. This time scale, TT(BIPM), is a realization of Terrestrial Time as defined by the International Astronomical Union (IAU).

\subsection*{Estimation of the error due to the intermittent operation of the OFS}
The performance of a time scale generated by intermittent optical steering is affected by the prediction error of the HM frequency. We estimated the magnitude of the error from the following analysis by using a simple noise model. The predictable trend of the HM frequency is expressed by the offset and the linear drift rate $d$ of the frequency, whereas the residual fluctuation is characterized by the Hadamard deviation. Here, we employ a model in which the prediction errors of the HM phase between two temporally separated OFS operations comprise two factors, which are the prediction error of the linear trend of the frequency and the stochastically accumulated phase error \cite{HM}. In the following estimation, we set the conditions that the HM frequency is regularly calibrated with a temporal separation of $\Delta T$ and that $N+1$ samples with fractional frequency $y_i (i=0\ldots N)$ are obtained in the latest duration $T=N\Delta T$ for drift estimation. We consider that the stochastic noise and residual white frequency noise cause a deviation $\Delta y_i$ from the linear  trend. The dominant noises of the OFS operation for $10^4$ s are flicker frequency noise and white frequency noise. The least-squares fitting to $\Delta y_i$ as a function of event time $\tau_i=i\Delta T-T/2$ gives a prediction error $\delta d$ of the linear drift rate of

\begin{equation}
\delta d = \frac{12}{\left( \Delta T\right) ^2 N (N+1) (N+2)}\sum^{N}_{i=0} \tau_i \Delta y_i.
\end{equation}

The error of the offset in the linear regression is the mean of $\Delta y_i \left( =\overline{\Delta y}\right) $ as $\tau_i$ is distributed symmetrically around $\tau =0$, which turns out to be $\overline{\Delta y} = (N+1)^{-1} \sum^{N}_{i=0}\Delta y_i$. Thus, the phase error $\Delta \phi_{\rm p}$ in the next free-running due to the prediction error is written as
\begin{equation}
\Delta \phi_{\rm p} = \int_{T/2}^{T/2+\Delta T} \left( \delta d \cdot t +\overline{\Delta y}\right) dt = \frac{6}{N(N+2)} \sum^{N}_{i=0} \tau_i \Delta y_i + \frac{T}{N(N+1)} \sum^{N}_{i=0} \Delta y_i .
\end{equation}
Here, we assume the normal distribution of $\Delta y_i$ with a standard deviation of $\sigma_{\rm p}$. Then, the accumulated phase error $\epsilon_{\rm p}$ due to the prediction error of the linear trend in $\Delta T$ is calculated as
\begin{equation}
\epsilon_{\rm p}\equiv \sqrt{E\left[ \left( \Delta \phi_{\rm p}\right)^2 \right]}=\frac{T}{N}\left[ \frac{(2N+1)(2N+3)}{N(N+1)(N+2)}\sigma_{\rm p}^2\right] ^{1/2},
\end{equation}
where $E\left[\left( \Delta \phi_{\rm p}\right)^2\right]$ is the expectation of $\left( \Delta \phi_{\rm p}\right)^2$.

The other factor is the accumulated stochastic phase error $\epsilon_{F}$ in the duration $T/2<t<T/2+\Delta T$, which is $\epsilon_{\rm F}^2=
\left( \sigma_{\rm F}\cdot \Delta T\right)^2 / \ln 2\ $ \cite{Allan} in the case that the flicker frequency noise $\sigma_{\rm H}\textbf{}(\tau )=\sigma_{\rm F}$ is predominant.  Thus, $E\left[|\Delta \phi|\right]$, the expected phase error between two OFS operations, is
\begin{equation}
E\left[|\Delta \phi|\right] = \left( \epsilon_{\rm p}^2+\epsilon_{\rm F}^2\right) ^{1/2} = \frac{T}{N}\left[\frac{(2N+1)(2N+3)}{N(N+1)(N+2)}\sigma_{\rm p}^2+\frac{1}{\ln 2}\sigma_{\rm F}^2 \right]^{1/2}.
\end{equation}

\subsection*{Uncertainty in TAI calibration}

The total uncertainty in the estimation of the one-month mean scale interval of TAI, which is indicated as error bars in Fig. 7, was determined by considering the following contributions. They are classified into factors attributed to (i) the atomic frequency standard, (ii) the link between the frequency standard and TAI, and (iii) the standard frequency as a secondary representation of the second. The small up-time ratio here imposes a link uncertainty larger than the uncertainty owing to the atomic frequency standard. The details of the contributions are described as follows.

\subsubsection*{Atomic frequency standard}
The uncertainty attributed to the Sr system comprises the type-A uncertainty $u_{\rm a}$ and the type-B uncertainty $u_{\rm b}$. $u_{\rm a}$ corresponds to the statistical uncertainty, which will be smaller if the measurement is performed for a longer time, whereas $u_{\rm b}$ corresponds to the systematic uncertainty of the standard. For the TAI evaluation of the one-month mean, the one-month campaign includes five or six operations with $10^4$ s OFS operations each. Since we evaluated the instability of NICT-Sr1 to be $7\times 10^{-15}/\tau^{1/2}$ by comparing two alternative servos \cite{Ca}, the estimated $u_{\rm a}$ for the total interrogation time of $\tau =5\times 10^4$ s is $3.1\times 10^{-17}$. The type-B uncertainty, in other words, the systematic uncertainty of NICT-Sr1 slightly varied during the campaign as described above. The mean with weights proportional to operation time was $6.9\times 10^{-17}$. Note that this uncertainty includes the uncertainty of the gravity shift, which amounts to $2.2\times 10^{-17}$.

\subsubsection*{Link}
The time-link uncertainty comprises that in the laboratory $u_{\rm l/Lab}$ and that for the satellite link $u_{\rm l/Tai}$ following the notation adopted in Circular T.
In the intermittent evaluation scheme proposed here, a major part of $u_{\rm l/Lab}$ is the possible uncertainty in the estimation of the one-month mean frequency of the HM based on the limited number of OFS operations. The deviation of the HM frequency from the linear trend is caused by the stochastic phase fluctuation and residual white noise of the HM. The HM frequency after the removal of the linear drift shows a flicker noise floor for an averaging time of one day to three weeks as shown in Fig. 1. We performed five or six homogeneously distributed measurements with roughly one-week separation for the linear fitting, in which the one-month mean frequency is obtained as the intercept in the fitting.

The expected error of the linear fitting based on the five or six $(=N+1)$ OFS calibrations is derived as follows. As discussed above, it is assumed that the measurements of the HM frequency deviate from the linear trend by $\Delta y_i$. Then, using the error of the resultant linear drift $\delta d\cdot t + \overline{\Delta y}$, the cumulative phase error in $-T/2<t<T/2$ due to the fitting error is
\begin{equation}
\delta \Phi \left( |t| < T/2\right) = \int^{T/2}_{-T/2} \left(\delta d\cdot t + \overline{\Delta y}\right)\ dt=\overline{\Delta y}T=\frac{T}{N+1}\sum^N_{i=0}
\Delta y_i.
\end{equation}

Therefore, assuming a normal distribution of $\Delta y_i$ with a standard deviation of $\sigma_{\rm p}=4\times 10^{-16}$, the standard deviation of $\delta \Phi$ turns out to be $\sigma_{\rm p}T/(N+1)^{1/2}$, which corresponds to a one-month mean frequency difference of $\sigma_{\rm p}/(N+1)^{1/2}=1.8\times 10^{-16}$ in the case of $N=4$. The normal distribution of $\Delta y_i$ assumed here probably underestimates the error as flicker noise is dominant. It is worth comparing this estimated uncertainty with the standard error of the linear fitting performed to determine the predictable drift of the HM frequency. The standard error of the fitting was typically $2.6\times 10^{-16}$, which is clearly larger than $\sigma_{\rm p}/(N+1)^{1/2}$. Considering these issues, the uncertainty $u_{\rm l/HMtrend}$ in the estimation of the linear drift is conservatively determined to be $2.6\times 10^{-16}$.
In terms of the effects of the stochastic phase fluctuation of the HM, the induced phase noise in a dead time of $\tau$ is known to be $\tau \sigma_{\rm H} /(\ln 2)^{1/2}$ in the case of flicker frequency noise \cite{Allan}. Thus, a non-operation time of one week results in a phase noise of 0.22 ns with $\sigma_{\rm H}=3\times 10^{-16}$. Since $30 / 7$ periods of 7 days are contained in one month, the uncertainty of the stochastic part $u_{\rm l/HMstoch}$ in the one-month mean frequency is estimated to be $0.22 \times 10^{-9} \times \sqrt{30/7}/(86400\times 30)=1.8\times 10^{-16}$.

Another minor part of $u_{\rm l/Lab}$ is the uncertainty of the DMTD system $u_{\rm l/DMTD}$. The uncertainty in five OFS operations amounts to $1\times 10^{-16}/\sqrt{5} = 4.5\times 10^{-17}$ as the noise of the DMTD system is $1\times 10^{-16}$ over an averaging time of $10^4$ s \cite{NICT:DMTD}. Finally, the link uncertainty in a local laboratory $u_{\rm l/Lab}$ is $u_{\rm l/Lab} =( {u_{\rm l/HMtrend}}^2+{u_{\rm l/HMstoch}}^2+{u_{\rm l/DMTD}}^2)^{1/2}=3.2\times 10^{-16}$.

The UTC-UTC(NICT) link uncertainty of 0.21 ns which we estimated in Ref. \citenum{NICT:OE}
corresponds to the uncertainty in the 5-day mean frequency of $u_{\rm 5d} =(\sqrt{2}\times 0.21\times 10^{-9})/(5\times 86400)=6.9\times 10^{-16}$ \ \cite{NICT:OE}. For a longer averaging time $t_{\rm d}$ in days, it is known that the uncertainty decreases to $u_{\rm 5d} (t_{\rm d} /5)^{-0.9}$ \ \cite{PanfiloLink}. Thus, the fractional uncertainty in the one-month mean frequency is $1.4\times 10^{-16}$.

\subsubsection*{Uncertainty in standard frequency}
The calibration of frequency with respect to a reference transition provided by non-Cs atoms has an additional uncertainty owing to the absolute frequency of the reference transition. For the case of the $^{87}$Sr clock transition, five absolute frequency measurements \cite{PTB:Optica,NICT:OE,NICT:APB,SYRTESr} were reported after the CIPM recommended frequency was updated in 2016 \cite{BIPM:standardF}. All these fifteen results give a weighted mean frequency of $\nu_{\rm mean}= 429\ 228\ 004\ 229\ 873.04(4)$ Hz with the fractional standard deviation of the mean being $u_{\rm SDOM}=9.1\times 10^{-17}$. Note that the scattering of the data is presumably determined by the reproducibility of the SI second and frequency links.

Finally, the uncertainty without $u_{\rm SDOM}$, namely $u_{\rm sub}=({u_{\rm a}}^2+{u_{\rm b}}^2+{u_{\rm l/Lab}}^2+{u_{\rm l/Tai}}^2)^{1/2}$, is $3.6\times 10^{-16}$. This uncertainty is at the same level as those obtained by the state-of-the-art PFSs. By incorporating the uncertainty of the standard frequency $u_{\rm SDOM}$, the total uncertainty for the TAI calibration proposed here amounts to $3.7\times 10^{-16}$.

%\subsection*{Data availability}
%The datasets generated during and/or analysed during the current study are available from the corresponding author on reasonable request.

\bibliography{sample}

\section*{Acknowledgements}

The authors thank G. Petit, T. Gotoh, H. Ito, M. Kumagai and M. Hosokawa for the fruitful discussions. Maintaining the operation of the lattice clock for half a year was achieved with the technical help provided by H. Ishijima and S. Ito.

%\section*{Author contributions}

%H. H. operated the lattice clock and calibrated the HM frequency with occasional help provided by T. I. F. N. processed the prediction of HM frequency with its implementation advised by Y. H. All authors extensively  discussed the planning, analyzed the result, and wrote the manuscript.

%\section*{Competing interests}
%The authors declare no competing interests.

\end{document}